\documentclass[prb,amsmath,amssymb,11pt,nofootinbib,showpacs,superscriptaddress]{revtex4}
\normalsize
\def\pnlabel#1{\hbox{\tiny[#1]}\label{#1}}
\newcommand{\cut}[1]{\error} %shouldn't be any left

\usepackage{graphicx}
\usepackage{bm}% bold math

\def\bead{_{\rm bead}}
\def\drift{_{\rm drift}}

\newcommand{\eem}[1]{\cdot10^{-#1}}

\def\sref#1{Sect.~\ref{#1}}
\def\fref#1{Fig.~\ref{#1}}
\def\frefn#1#2{Fig.~\ref{#1}{\it #2}}

\newcommand{\inv}{^{\raise.15ex\hbox{${\scriptscriptstyle -}$}\kern-.05em 1}}

\newcommand{\eff}{_{\rm eff}}

\def\kbt{{k_{\rm B}T}}
\def\tot{_{\mathrm{tot}}}

\newcommand{\pitemize}{\begin{itemize}\setlength{\itemsep}{0pt}\setlength{\parsep}{0pt}}
\newcommand{\xpitemize}{\end{itemize}}
\newcommand{\penumerate}{\begin{enumerate}\setlength{\itemsep}{0pt}\setlength{\parsep}{0pt}}
\newcommand{\xpenumerate}{\end{enumerate}}

\def\ifigfull#1#2#3#4{\begin{figure}%[tb!]
\begin{center}\includegraphics{#3}
\end{center}\smallskip 
\caption{\footnotesize #2 \pnlabel{#1}}
\end{figure}}

\begin{document}

\title{Elementary simulation of tethered Brownian motion}
\author{John F. Beausang}  \affiliation{Department of
Physics and Astronomy, University of Pennsylvania, Philadelphia PA
19104, USA}
\author{Chiara Zurla}
\affiliation{Department of Physics, Emory University, Atlanta GA 30322}
\author{Luke Sullivan}
\affiliation{Ursinus College, Collegeville, PA 19426-1000}
\author{Laura Finzi}
\affiliation{Department of Physics, Emory University, Atlanta GA 30322}
\author{Philip C. Nelson}
\email{nelson@physics.upenn.edu} \affiliation{Department of
Physics and Astronomy, University of Pennsylvania, Philadelphia PA
19104, USA}

\date{\today}
\begin{abstract}
We describe a simple numerical simulation, suitable for an
undergraduate project (or graduate problem set), of the Brownian motion
of a particle in a Hooke-law potential well. Understanding this
physical situation is a practical necessity in many experimental
contexts, for instance in single molecule biophysics; and its
simulation helps the student to appreciate the dynamical character of
thermal equilibrium. We show that the simulation succeeds in capturing
behavior seen in experimental data on tethered particle motion.

%DO: Shall we mention that theory says $C(\tau)\propto\exp(-\kappa|\tau|/\zeta)$?

\end{abstract}

\maketitle

\section{Introduction and summary}
Introductory courses in statistical physics often place their greatest
emphasis on average quantities measured in thermodynamic equilibrium. Indeed, the study of equilibrium
gives us many powerful results without having to delve into too much 
technical detail. This simplicity stems in 
part from the fact that for such problems, we are not interested in time 
dependence (dynamics); accordingly dissipation constants like 
friction and viscosity do not enter the formulas.

There are compelling reasons, however, to introduce
students to genuinely dynamical aspects of thermal systems as early as 
possible---perhaps even before embarking on a detailed study of 
equilibrium \cite{philbook}. One reason is that 
students can easily miss the crucial steps needed to go from a basic
appreciation that ``heat is motion'' to understanding the Boltzmann
distribution, and thus can end up with a blind spot in their understanding of 
the foundations of the subject. Although any kind of rigorous proof of 
this connection is out of place in a first course, nevertheless a 
demonstration of how it works in a sample calculation can cement the 
connection.

A second reason to give extra attention to dynamical phenomena is the
current surge in student interest in biological physics. Much current
experimental work studies the molecular processes of life, or their
analogs, at the single-molecule level, where simple mathematical
descriptions do seem to capture the observed behavior.

One familiar setting where simple models describe statistical dynamics 
well is the theory of the random walk, and its relation to diffusion. 
Ref.~\cite{philbook} showed an attempt to present classical statistical 
mechanics starting from the random walk, building on earlier classics 
such as Ref.~\cite{Bberg93a}. The link between thermal 
motion and the Boltzmann distribution emerges naturally in the 
analysis of sedimentation equilibrium: We require that in 
equilibrium, diffusive changes in a concentration distribution must 
cancel changes caused by drift from a constant external force field (gravity).

In this article we discuss a generalization of free Brownian motion
that is important for interpreting a large class of current experiments
in single-molecule biophysics: The Brownian motion of a particle
tethered to a Hookean spring. Specifically, we investigate the 
dynamics of fluctuations of such a particle in equilibrium.
Although the theoretical analysis of this
problem does appear in some undergraduate textbooks, it sometimes appears
forbiddingly complex (e.g. Ref.~\cite{Breif65a}). We have found, however,
that numerical simulation of the system is well within the range of an
undergraduate project. Either as a project or a classroom 
demonstration, such a simulation brings insight 
into the emergence of equilibrium behavior from independent random 
steps, and also can serve as an entry into the topic of equilibrium 
fluctuations. 

The appendix gives an implementation written by one of us, an 
undergraduate at Ursinus College, in \textsc{matlab}.

\section{Experimental background}
As motivation, we briefly mention two contexts in which Brownian 
motion in a harmonic (Hooke-law) trap has played a role in recent 
biological physics experiments.

Optical trapping\cite{gitt97a,philbook} is now an everyday tool for the manipulation of 
micrometer-scale objects (typically a polystyrene bead), and 
indirectly of nanometer-scale objects attached to them (typically DNA, 
RNA, or a protein). In this method, a tightly focused 
laser spot creates a 
restoring force tending to push a bead toward a particular point in space.
When the trapping beam has a Gaussian profile, the 
resulting force on the bead is often to a good approximation linear in 
bead displacement. Thus the bead executes Brownian motion in a 
harmonic potential well. In such a well the motions along the three
principal axes of the well are independent. 

The bead's motion in one or two dimensions can be tracked to high
precision, for example by using interferomotry, thus yielding a time
series. The probability distribution of the observed bead locations
then reflects a compromise between the restoring force, pushing the 
bead to the origin, and thermal motion, tending to randomize its 
location. The outcome of this compromise is a Gaussian distribution of 
positions, from which we can read
the strength of the harmonic restoring force (``trap
stiffness''). For practical reasons, however, it is often more accurate to
obtain both trap stiffness and the bead's effective friction constant
from the autocorrelation function of the bead position (see
 \sref{s:sr} below). For example, slow microscope
drift can spoil the observed probability distribution function (see
Ref.~\cite{gitt97a}).

Our second example concerns tethered particle motion\cite{finzi95}. In this
technique, a bead is physically attached to a ``tether'' consisting of 
a long single strand of DNA. The other end of the tether is anchored to
a microscope slide and the resulting bead motion is observed. Changes
in the bead's motion then reflect conformational changes in the tether,
for example the binding of proteins to the DNA or the formation of a
long-lived looped state. \frefn{f:tpm}a shows example data for a
situation where such conformational changes are absent, that is, simple
tethered particle motion.

\ifigfull{f:tpm}{\textit{Left,} sample experimental data for the
$x$-component of the motion of a bead 
of radius $R\bead\approx240\,$nm, attached to a DNA tether of length 
$L_{\rm tether}\approx3500$ basepairs, or
$\approx1200\,$nm. The experiment, and the protocols used to remove 
drift from the raw data, are described in detail in Ref.~\cite{nels06a}. For clarity only the first 
200$\,$s of data are shown; the full data run lasted 600$\,$s.
\textit{Right,} logarithm of the autocorrelation function of 
$x$ expressed in nm$^2$ 
(see text). \textit{Solid line,} experimental data. \textit{Dots,} 
simulation described in this paper, using parameters $A\eff=72\,$nm and 
effective viscosity 2.4 times that of water in bulk.
}{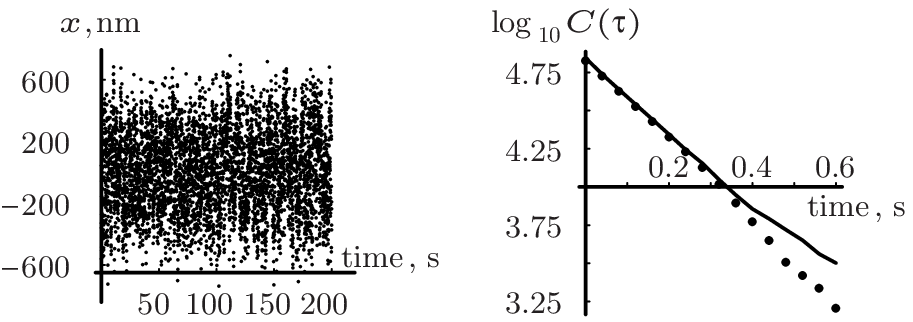}{}

\ifigfull{f:histo}{\textit{Left,} histogram of measured bead $x$ 
position for the experimental data shown in \fref{f:tpm}. The solid 
curve shows a Gaussian distribution with the same normalization and 
variance.
\textit{Right,} similar histogram for our simulation.}{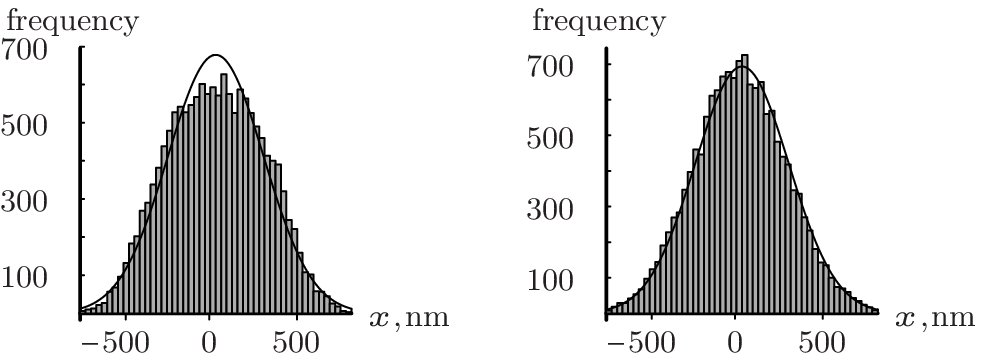}{}

As in the optical trap case, one can discard the dynamical information
in the time series by making a histogram of particle locations.
\frefn{f:histo}a shows the frequencies of occurence of various values of 
$x$. Rather detailed agreement between theory and experiment has been
obtained for these histograms, including the slight deviation from 
Gaussian distribution shown in the figure\cite{sega06a,nels06a}. In the present
note, however, we are interested in a less sophisticated treatment of a
more general question:  Can we understand at least some aspects of the
\textit{dynamical} information contained in data like those in \fref{f:tpm}?

To this end,
\frefn{f:tpm}b shows the logarithm of the autocorrelation 
function, $C(\tau)\equiv\langle x(t+\tau)x(t)\rangle$, where the
brackets denote averaging over $t$. At $\tau=0$ this quantity is just
the meansquare displacement, which would diverge for a free particle
but instead has a finite value determined by the equipartition theorem. At large
times the autocorrelation falls to zero, because two independent
measurements of $x$ are as likely to lie on opposite sides of the
tethering point as they are to lie on the same side. In fact, the
autocorrelation function should fall exponentially with $\tau$, as we 
see it does in \frefn{f:tpm}b.\cite{Breif65a}

\section{Simulation setup}
We wish to simulate the motion of a bead of radius $R\bead$, attached
to a tether of length $L\tot$, and compare
our results to experimental data. For this we will need to know a 
specific property of DNA
in typical solution: Its
``persistence length'' $A$ is $A\approx 45\,$nm.\cite{philbook}

We will suppose that the external forces acting on the bead are a
hard-wall repulsion from the microscope slide, a tension force from the tether,
and random collisions with surrounding water molecules (see \sref{s:d}
for further discussion). The tension force produces an effective
potential well that keeps the bead close to its attachment point. In
fact, at low relative extension, the tension exerted by a semiflexible
polymer like DNA is approximately given by a Hooke-type
law:\cite{philbook} $f=-\kappa x$. The effective spring constant is
$\kappa=3\kbt/(2AL\tot)$, where $\kbt\approx4.1\eem{21}\,$J is the
thermal energy at room temperature
%$A\approx50\,$nm is the persistence length of DNA,
and $L\tot$ is the contour length of the polymer. Thus, again the
motions in each of the $x$, $y$, and $z$ directions are independent.
Because the microscopy observes only $x$ and $y$, we can reduce the problem to a two-dimensional 
one, and hence forget about the hard-wall force, which acts only along 
$z$. In fact, we will reduce it still further, by examining only the
$x$ coordinate of the bead position.

There is a subtlety in that we do not directly observe the endpoint of 
the polymer in an experiment. Rather, we observe the image of the 
bead; the image analysis software reports the location of the bead 
center, a distance $R\bead$ from the attachment point. Thus the time 
series in \frefn{f:tpm}a reflects the motion of the endpoint of a 
composite object, a semiflexible polymer attached by a flexible link 
to a final, stiff segment of length $R\bead$. To deal simply with this 
complication, we note that a semiflexible polymer 
can also be approximately regarded, for the purposes of finding its 
force--extension relation, as a chain of stiff segments of length 
$2A$. In our case $2A\approx100\,$nm is not much larger than
$R\bead$, so we will approximate the system as a single Hookean 
spring, with effective length $L\tot=L_{\rm tether}+R\bead$ and an 
effective persistence length $A\eff$ slightly larger than $A$. In the 
data we present, $L_{\rm tether}\approx3500$ basepairs, or
$\approx1200\,$nm. We will fit the data to obtain $A\eff$.

Suppose we observe the bead at time intervals of $\Delta t$. Without
the tether, the bead would take independent random steps, each a
displacement drawn from a Gaussian distribution with mean-square step
length $2D(\Delta t)$, where $D$ is the bead's diffusion constant. If
the bead were subjected to a \textit{constant} force $f$ (for example gravity),
we could get its net motion by superimposing an additional
deterministic drift on the random steps: $ \Delta\drift x=f/\zeta$. The
friction constant $\zeta$ is related to  $D$ by
the Einstein and Stokes relations: $\zeta=\kbt/D=6\pi\eta R\bead$,
where $\eta$ is the viscosity of water. For the tethered case, at each
step we instead use a position-dependent force $-\kappa x$, where $x$ is the
current displacement. For small enough $\Delta t$ (perhaps smaller than
the actual video frame rate), $x$ will be roughly constant during the
step, justifying this substitution.

Here again we find a subtlety: The presence of the nearby wall creates 
additional hydrodynamic drag on the bead \cite{happ83a,scha06a}. Moreover, the tether itself 
incurs significant hydrodynamic drag impeding the system's motion. 
Again for simplicity, we acknowledge these complications by fitting to 
obtain an effective viscosity $\eta\eff$, which we expect to be 
larger than the value $10^{-3}\,$Pa$\,$s appropriate to water in bulk.

\section{Simulation results\pnlabel{s:sr}}
To summarize, the simulation implements a Markov process. Each step is 
the sum of a random, diffusive component with meansquare  $2D(\Delta 
t)$, and a drift component $-D\kappa x/\kbt$. The constants 
$D=\kbt/(6\pi\eta\eff R\bead)$ and 
$\kappa=3\kbt/(2A\eff L\tot)$ contain two unknown fit parameters, the effective persistence 
length $A\eff$ and viscosity $\eta\eff$. The output of the simulation 
is the probability distribution of positions, and the autocorrelation 
function, which may be compared to experimental data.

The simulation is deemed successful to the extent that the two fit
parameters take values reasonably close to the expected values,
differing in the expected directions, and the full functional forms of
the outputs agree with experimental data. Figures \frefn{f:tpm}b and
\fref{f:histo} show that indeed the simple model works well. Our 
simulation took $\Delta t=0.625\,$ms, for a total of about a million 
steps, which were then sampled every 40$\,$ms to resemble the 
experimental data.

\section{Discussion\pnlabel{s:d}}
Our mathematical model made some naive simplifications. Two which have 
been mentioned involve the role of the bead radius, and certain 
sources of drag on the bead. In addition, there is a time scale for 
rearrangements of the DNA needed to change its extension, and for 
rotatory diffusion of the bead, which changes the location of the 
attachment point relative to its center. All of these effects have 
been assumed to be lumped into effective values of the fit parameters.

Despite these simplifications,
however, we did obtain two key qualitative aspects of the experimental 
data as outputs from the model, not set by hand. First, the autocorrelation
function of equilibrium fluctuations has the expected exponential form. Moreover,
as a consistency check, the experimental histogram of bead positions 
had roughly the
Gaussian form we would expect for motion in a Hookean potential well.
Both of these results emerge as statistical properties of a large
number of simple steps, each involving only a diffusive step combined
with a drift step based on the current bead location.

The insights obtained from this exercise are different from those obtained from the 
analytical solution; for example, students can see the average 
behavior emerging from the random noise as the simulation size grows.
In addition, the simulation approach opens the door to replacing the 
assumption of a harmonic potential by any other functional form. 
% Thus 
% a student could improve the simulation by reading a corrected 
% potential $U(x)$ directly from \frefn{f:histo}a, feeding
% it into the simulation, and then fitting the autocorrelation function by 
% adjusting the value of $\eta\eff$. An anharmonic choice of $U$ would, 
% however, invalidate the independence of xyz...

\begin{acknowledgments}
We thank I. Kulic and R. Phillips for many discussions. 

This work was partially supported by the Human Frontier Science
Programme Organization (LF and PN), and the National Science Foundation under Grant DMR-0404674
and the NSF-funded NSEC on Molecular Function at the Nano/Bio Interface
DMR04-25780 (PN). LF and PN acknowledge the hospitality of the Kavli
Institute for Theoretical Physics, supported in part by the National
Science Foundation under Grant PHY99-07949.
\end{acknowledgments}

\bibliographystyle{prsty}%unsrt}
\bibliography{tethered}
\appendix
\section{Matlab code}
\small
\begin{verbatim}
function TetheredParticleAnalysis (Xdata,Ncorr,Nbins,deltat)
%This function simulates and analyzes the motion of a 1D random walk
%confined in a harmonic potential well.
%
%Written by Luke Sullivan, Ursinus College
%Edited by John F. Beausang, University of Pennsylvania
%
%Xdata    = array of bead position (nm)
%nCorr    = number of points in correlation function
%Nbins    = number of histogram bins
%deltat   = time step of data and simulation (sec)

%%%%Experimental Data%%%%
Xdata     = transpose(Xdata); 
n         = length(Xdata);    %number of data points
time      = (1:n)*deltat;     %time series
[FData,rData,histoData] = GaussHistoX(Xdata,Nbins);
logACData = LogAutoCorr(Xdata,Ncorr,deltat);

%%%%Simulated data%%%%
Xsim = RandWalkSim(length(Xdata),deltat);  %generate simulated data
[FSim,rSim,histoSim] = GaussHistoX(Xsim,Nbins);
logACSim = LogAutoCorr (Xsim,Ncorr,deltat);

%%%%output%%%%
subplot(2,3,1);plot(time,Xdata,'b')%Experimental data
title ('Experimental data');
xlabel ('Time (sec)');
ylabel ('Bead Position (nm)');
subplot(2,3,2);plot(rData,FData,'bx',histoData(1,:),histoData(2,:),'bo')%Gaussian curve
title ('Gaussian fit (x) and Histogram (o) of exp data');
xlabel ('position (nm)');
ylabel ('Probability (1/nm)');
subplot(2,3,4);plot(time,Xsim,'r')%Simulated data
title ('Simulated data');
xlabel ('Time (sec)');
ylabel ('Bead Position (nm)');
subplot(2,3,5);plot(rSim,FSim,'rx',histoSim(1,:),histoSim(2,:),'ro')%Gaussian curve
title ('Gaussian fit (x) and Histogram (o)of Sim data');
xlabel ('position (nm)');
ylabel ('Probability (1/nm)');
subplot(2,3,3);plot(logACData(1,:),logACData(2,:),'b-',logACSim(1,:),logACSim(2,:),'r-')%plot of correlation function of data
title ('Compare AutoCorrelation');
xlabel ('time difference (sec)')
ylabel ('Log[AutoCorrelation], (nm2)')

%%%%Subroutines%%%%%

function [F,r,Xhisto]=GaussHistoX (Xdata,Nbins)
%This function histograms the data and fits a Gaussian distribution
Xmax        = max(abs(Xdata));  %maximum position
n           = length(Xdata);    %number of data points
binWidth    = Xmax/Nbins;       %histogram bin width
stdevX      = std(Xdata);       %standard deviation of data
F=zeros(1,n);r=zeros(1,n);Xhisto=zeros(2,Nbins+1);  %initialize
Xhisto(1,:)= ((1:Nbins+1)-.5)*binWidth;     %midpoint of histogram bins
for i=1:n
    r(i)=abs(Xdata(i));
    F(i)=2/sqrt(2*pi*stdevX^2)*exp(-(Xdata(i)^2)/(2*stdevX^2));%1 sided gaussian curve
    which=1+floor(abs(Xdata(i))/binWidth);  %which bin data falls into
    Xhisto(2,which)=Xhisto(2,which)+1;      %increment bin
end
Xhisto(2,:)=Xhisto(2,:)/n/binWidth; %convert counts to probability (1/nm)

function logac = LogAutoCorr (Xdata,Ncorr,deltat)
%This function determines the autocorrelation of the data for Ncorr points
n         = length(Xdata);
logac     = zeros(2,Ncorr);
logac(1,:)= (0:Ncorr-1)*deltat;      %time steps
for s = 1:Ncorr
    temp = zeros (1,n-s+1);
    for i=1:(n-s+1)
        temp(i)=Xdata(i)*Xdata(i+s-1);
    end
    logac(2,s)=log10(sum(temp)/(n-s+1));
end

function Xsim=RandWalkSim(n,deltat)
%This function simulates a 1D random walk in a harmonic potential
%%%%physical parameters%%%%
L       = 3477*.34;         %tether length (nm)
Xi      = 72;               %tether persistence length (nm)
Rb      = 240;              %bead radius (nm)
kbT     = 4.1*10^(-21);     %thermal energy (J)
eta     = 2.4*10^(-30);     %viscosity of H2O (J*s/nm^3)
D       = kbT/(6*pi*eta*Rb);%Stokes diffusion constant (nm2/s)
kappa   = 3/2*kbT/Xi/L;     %spring constant (J/nm2)
mu      = D*kappa/kbT*deltat;%bias of step in harmonic potential
ldiff   = sqrt(2*D*deltat); %diffusion length (nm)

Xsim=zeros(1,n);
for i=2:n
    deltaX=randn(1)*ldiff-Xsim(i-1)*mu;%step size for each element called
    Xsim(i)=Xsim(i-1)+deltaX;%new element value
end
\end{verbatim}
\end{document}